\documentclass[aps,prl,superscriptaddress,twocolumn,floatfix,nofootinbib,preprintnumbers,showpacs]{revtex4}
\usepackage{amsmath,multirow,graphicx,tabularx,epsfig}

\def\pveto{P_\text{veto}}
\def\nj{n_\text{jets}}

\newcommand\one{\leavevmode\hbox{\small1\normalsize\kern-.33em1}}

\newcommand{\qqquad}{\qquad \qquad}

\newcommand{\matx}{|\mathcal{M}|^2}
\newcommand{\gev}{{\ensuremath\rm GeV}}

\def\slashchar#1{\setbox0=\hbox{$#1$}           % set a box for #1
   \dimen0=\wd0                                 % and get its size
   \setbox1=\hbox{/} \dimen1=\wd1               % get size of /
   \ifdim\dimen0>\dimen1                        % #1 is bigger
      \rlap{\hbox to \dimen0{\hfil/\hfil}}      % so center / in box
      #1                                        % and print #1
   \else                                        % / is bigger
      \rlap{\hbox to \dimen1{\hfil$#1$\hfil}}   % so center #1
      /                                         % and print /
   \fi}

\def\eg{{\sl e.g.} \,}
\def\ie{{\sl i.e.} \,}

\newcommand{\be}{\begin{eqnarray*}}
\newcommand{\ee}{\end{eqnarray*}}

\newcommand{\bee}{\begin{eqnarray}}
\newcommand{\eee}{\end{eqnarray}}
\newcommand{\beeq}{\begin{equation}}
\newcommand{\eeeq}{\end{equation}}

\begin{document}


\date{\today}

\title{Understanding Jet Scaling and Jet Vetos in Higgs Searches}
% Jet Counting in Higgs Production

\author{Erik Gerwick}
%\email{e.gerwick@sms.ed.ac.uk}
\affiliation{SUPA, School of Physics \& Astronomy, The University of Edinburgh, 
             Scotland}

\author{Tilman Plehn}
%\email{plehn@uni-heidelberg.de}
\affiliation{Institut f\"ur Theoretische Physik, Universit\"at Heidelberg,
             Germany}

\author{Steffen Schumann}
%\email{s.schumann@thphys.uni-heidelberg.de}
\affiliation{Institut f\"ur Theoretische Physik, Universit\"at Heidelberg,
  Germany}
\affiliation{II. Physikalisches Institut, Universit\"at G\"ottingen, 
  Germany}

\begin{abstract}
  Jet counting and jet vetos are crucial analysis tools for many LHC
  searches.  We can understand their properties from the distribution
  of the exclusive number of jets. LHC processes tend to show either a
  distinct staircase scaling or a Poisson scaling, depending on
  kinematic cuts. We illustrate our approach in a detailed study of
  jets in weak boson fusion Higgs production.
\end{abstract}

\maketitle

%%%%%%%%%%%%%%%%%%%%%%%%%%%%%%%%%%%%%%%%%%%%%%%%%%%%%%%%%%%%%%%%%%%%%%%%

\underline{Higgs searches} --- at hadron colliders are typically
plagued by large backgrounds. To extract Higgs signals we develop
strategies to suppress backgrounds where the structure of QCD effects
plays an important role.

Searching for a Higgs boson produced we first reject backgrounds by
reconstructing the invariant or transverse mass of the Higgs decay
products. Based on the gluon-fusion production mechanism an additional
discriminating feature turns out to be the boost of the Higgs boson,
\ie its recoil against a number of quarks or gluons. If the Higgs boson is
usually produced with small transverse momentum we require a large
opening angle of its decay products and veto hard jet activity.

Combined Higgs search results based on exactly zero jets and one jet
have recently been shown for the LHC~\cite{atlas,cms}. For each case
the kinematic extraction cuts can be optimized once we understand how
signal and backgrounds fall into these jet bins.  

Extending these to two jets will include Higgs bosons produced in weak
boson fusion (WBF), \ie in association with two so-called tagging
jets~\cite{wbf}. For this process the color structure of the signal
together with the two n jets forbids gluon exchange between the
incoming quarks. Correspondingly, jet radiation into the central
detector is suppressed.  Requiring exactly two jets is equivalent to a
central jet veto as a means to suppress backgrounds~\cite{cjv}.

As another example, the production of a Higgs boson in association
with a $W$ or $Z$ boson~\cite{wh} is mediated by an electroweak
process. We therefore do not expect significant jet activity in the
signal, in contrast to $W/Z$+jets backgrounds. A jet veto makes use of
this feature.\medskip

In all these cases we need to compute and measure the numbers of jets
in signal and background channels. From the theory perspective the
requirement of observing exactly $\nj$ jets is
problematic~\cite{scet}. Any computation based on parton densities
which obey the DGLAP equation is {\sl per se} jet inclusive, \ie we
always compute cross sections including an unspecified number of
collinear jets. In contrast, a jet veto corresponds to the probability
\begin{equation}
1 - \pveto 
= \frac{\sigma_0}{\hat{\sigma}_\text{tot}} 
= \frac{\hat{\sigma}_\text{tot} - \hat{\sigma}_1}{\hat{\sigma}_\text{tot}} \; ,
\label{eq:prob}
\end{equation}
in terms of the inclusive ($\hat{\sigma}_n$) and exclusive
($\sigma_n$) $n$-jet-associated cross sections and $\hat{\sigma}_0
\equiv \hat{\sigma}_\text{tot}$. 

The problem with Eq.\eqref{eq:prob} is that we cannot really include a
sensitive detector region for the number of (visible) jets.  The
collinear limit of the DGLAP equation is not a useful approximation
because we only observe jets with finite transverse momentum. In an
improved description the initial state parton shower unfolds the
parton splittings and produces more or less collinear initial state
radiation. The approximate simulation of initial state radiation over
the entire phase space critically limits our understanding of the {\sl
  exclusive} $\nj$ distribution and associated veto survival
probabilities.\medskip

Matrix element and parton shower matching is the key to simulating jet
radiation at the LHC~\cite{early_matching,ckkw,mlm}.  It correctly 
describes the radiation of any number of jets over the entire
radiation phase space, for example for pure QCD jet events~\cite{us},
$W,Z$ or photon production in association with
jets~\cite{scaling_orig,ex_lhc,us,photon}, or Higgs production in gluon
fusion~\cite{LH_2009_ggH} as well as in weak boson fusion. 

We propose to study exclusive $\nj$ distributions. A jet veto then
becomes nothing but a cut on another well understood distribution.  It
will turn out that jet radiation is usually governed by one of two
patterns, {\sl staircase scaling} and {\sl Poisson scaling}. We have 
access to both of them with current LHC sample sizes.  Based on
such scaling studies we can for the first time count exclusive jets in
a completely testable framework.\medskip

\underline{Staircase scaling} --- is defined as constant ratios of
successive $\nj$ rates for example for $W$+jets or pure QCD jets
production
\begin{equation}
R_{(n+1)/n} = \frac{\sigma_{n+1}}{\sigma_n}
       \equiv R  \; . 
\label{eq:staircase1}
\end{equation}
The scaling parameter $R$ depends on the core process and on the
requirements on the jets, but not on $n$. This feature has been
observed at UA2, Tevatron and the LHC~\cite{ex_lhc,scaling_orig}.
Historically, Eq.\eqref{eq:staircase1} is defined inclusively as
$\hat{R}$. However, exclusive and inclusive staircase scaling is
equivalent with $R \equiv \hat{R}$~\cite{us}. Following the above
argument, we rely on the exclusive formulation to study the $\nj$
distribution.  From the constant jet ratio $R$ we can derive the
normalized distribution of the exclusive number of jets
\begin{equation}
\sigma_n = \sigma_0 \; e^{-bn} \qqquad \text{with} \quad
R \equiv e^{-b} \; .
\label{eq:staircase2}
\end{equation}
%

%--------------------------------------------------------
\begin{figure}[b]
\includegraphics[width=0.48\textwidth]{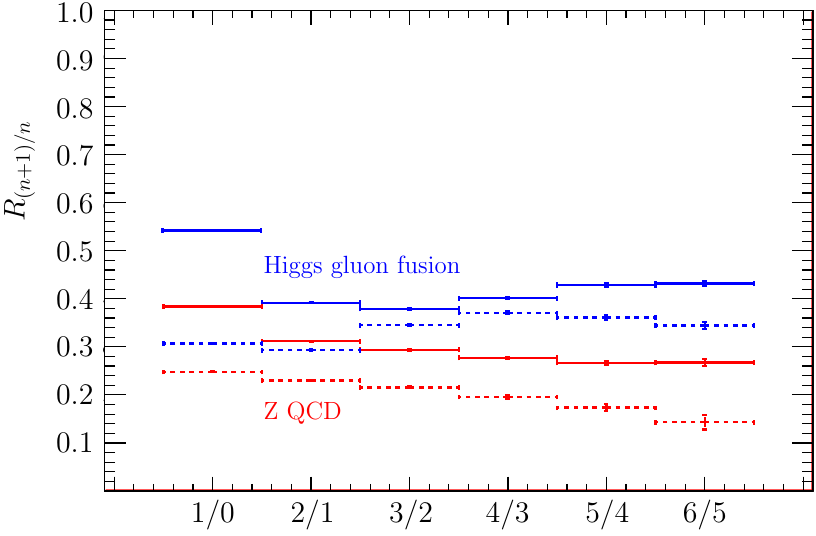}
\caption{Simulated $R_{(n+1)/n}$ distributions for $Z$+jets and
  $H$+jets production via the effective $g$-$g$-$H$ coupling. We only
  apply the basic cuts $p_{Tj}>20 (30)$~GeV and $|y_j|<4.5$ for the
  solid (dotted) entries. The quoted $n$ values are in addition to the
  two leading jets.}
\label{fig:staircase1}
\end{figure}
%--------------------------------------------------------

At the LHC we can study staircase scaling for example in $W$+jets and
$Z$+jets production~\cite{scaling_orig,us,ex_tev,ex_lhc}, in pure QCD
jet production~\cite{us}, and in $\gamma$+jets
production~\cite{photon}.  QCD production of $Z$+jets is defined by as
$Z$ radiation off strong jet production $\matx \propto \alpha
\alpha_s^n$.

To observe staircase scaling it is crucial to consider total cross
sections with as few kinematic cuts as possible. While the reason for
this scaling behavior from first principles is not entirely clear, we
know that it is closely linked with the non-Abelian nature of QCD. In
our simulations we see that the majority of jets arise through initial
state radiation (ISR) off the parton entering the hard process. This
radiation mediates between the virtuality scales of a parton inside
the proton and the parton entering the hard process. The large jet
multiplicities constituting staircase scaling are then driven by
further splitting of very few ISR quarks or gluons, \ie we are
sensitive to final state radiation patterns starting from hard
ISR. This splitting ISR feature is driven by multiple gluon splitting,
\ie it only occurs for non-Abelian massless gauge bosons with a self
coupling.\medskip

In Fig.~\ref{fig:staircase1} we see that with an approximately
constant $R$ {\sc Sherpa}~\cite{sherpa} based on {\sc Ckkw}
matching~\cite{ckkw} with at least four hard jets correctly reproduces
experimental observations. Note that the quoted values for $(n+1)/n$
are counted in addition to the two tagging jets which define our core
process.

A proper analysis of the theoretical uncertainties shows that this scaling
behavior is not affected~\cite{us}. Varying $\alpha_s(m_Z) = 0.114 -
0.122$~\cite{cteq10} merely shifts $R$. A consistent variation of all
renormalization, factorization and shower scales also leaves the
scaling feature untouched, but with a large enough shift in $R$ that
such a shift should be viewed as a tuning parameter for jet
merging~\cite{us}. Fixed-order QCD corrections leave the scaling
untouched~\cite{blackhat}. The only complication we observe in
Fig.~\ref{fig:staircase1} is the variation of the first entry
$R_{1/0}$ which corresponds to a need to properly define the hard
process. Counting only jets above 30~GeV instead of 20~GeV alleviates
this problem.\medskip

%--------------------------------------------------------
\begin{figure}[b]
\includegraphics[width=0.47\textwidth]{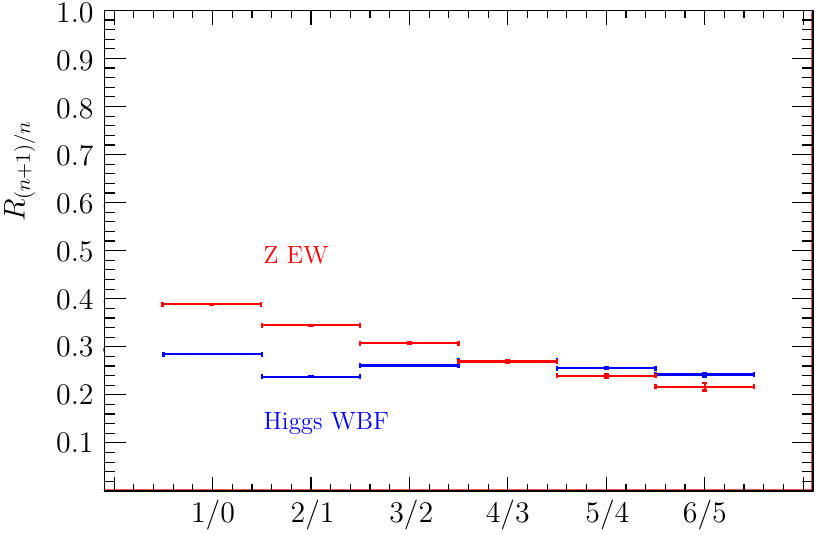}
\caption{$R_{(n+1)/n}$ distributions for electroweak $Z$+jets
  production and WBF Higgs production. Again, we only apply
  $p_{Tj}>20$~GeV and $|y_j|<4.5$ and count $n$ in addition to the
  two leading jets.}
\label{fig:staircase2}
\end{figure}
%--------------------------------------------------------

In Fig.~\ref{fig:staircase2} we show the $R_{(n+1)/n}$ distribution
for the WBF Higgs signal as well as the electroweak $Z$+jets channel 
at order $\matx \propto \alpha^3 \alpha_s^n$. From signal and 
background studies for WBF Higgs production~\cite{wbf,cjv} we know 
that these processes have different jet radiation patterns.  However, 
for the total cross section with minimal cuts we still observe an 
approximate staircase scaling $R_{(n+1)/n} =$~const. The slight drop 
in $R_{(n+1)/n}$ is due to different classes of Feynman diagrams
contributing to the electroweak process at different jet
multiplicities, including WBF topologies, $Z$ bremsstrahlung, and
$WW/WZ$ pair production\medskip

\begin{figure}[t]
\includegraphics[width=0.47\textwidth]{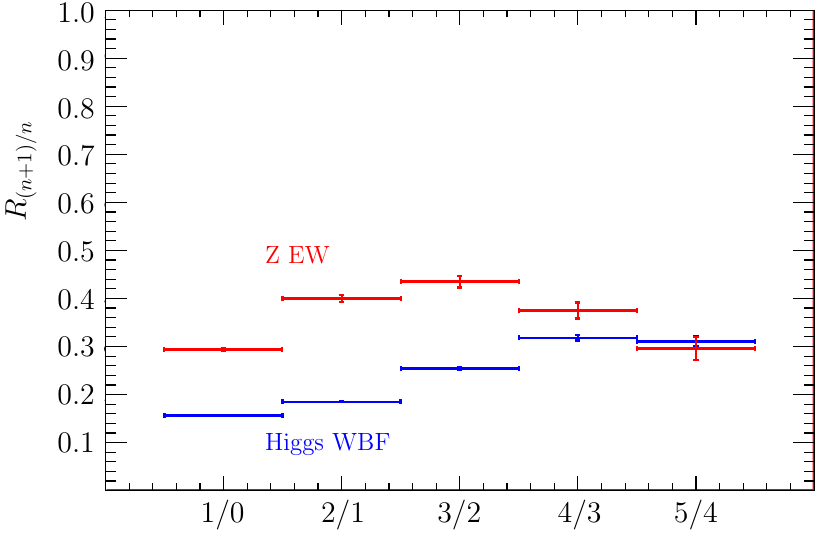}
\caption{$R_{(n+1)/n}$ distributions for electroweak $Z$+jets
  production, and Higgs production in weak boson fusion.  Unlike in
  Fig.~\ref{fig:staircase2} we now count only additional jets in the
  veto region defined by Eq.\eqref{eq:veto}.}
\label{fig:staircase3}
\end{figure}

If we want to use jet scaling for Higgs searches, we need to consider 
background suppression cuts. As an illustrative example we study
WBF Higgs production $qq \to qqH$ with a decay $H \to \tau \tau$ or $H
\to WW$~\cite{wbf}. Fixed-order corrections to the inclusive rate are
known to be moderate~\cite{wbf_nlo}. We are only interested in the
production process so neglect the Higgs decay products entirely.
Starting from exactly two tagging jets defined as the most forward  
and backward jets with
\begin{alignat}{5}
\label{eq:wbf}
p_{T,j} &> 20~\gev \qquad 
& |y_j| &< 4.5 \\
y_1 y_2 &< 0 \qquad 
& |y_1 - y_2| &> 4.4 \qquad 
m_{jj} > 600~\gev \notag 
\end{alignat}
we compute the jet activity in the to-be-vetoed region
\begin{equation}
p_T^\text{veto} > 20~\gev
\qquad 
\min y_{1,2} < y^\text{veto} < \max y_{1,2} \; .
\label{eq:veto}
\end{equation}
In Fig.~\ref{fig:staircase3} we first show the signal-like processes,
now only counting jets inside the veto region
Eq.\eqref{eq:veto}. These cuts reduce backgrounds while capturing the
signal properties, most notably the large invariant mass of the
tagging jets.  Therefore, they do not have a major effect on weak
boson fusion and electroweak $Z$ production and the approximate
staircase scaling persists. The suppression of the first additional
jets merely reflects the suppression due to the color structure
combined with the strictly enforced tagging jet structure.  Slight
deviations from a perfect staircase scaling are also expected because
the WBF cuts in Eq.\eqref{eq:wbf} are given by the LHC analyses and
not optimized to test scaling features.\medskip

%--------------------------------------------------------
%--------------------------------------------------------

\underline{Poisson scaling} --- is a different scaling property which
we observe for example in QCD $Z$+jets production after cuts. For a
limited number of emissions up to $n^\text{crit}$ it reads
\begin{equation}
R_{(n+1)/n}
= \frac{\sigma_{n+1}}{\sigma_n}  
= \frac{\bar{n}}{n+1} 
\qquad \text{for} \quad n < n^\text{crit}  \; ,
%\; \to \; \frac{\bar{n}}{n+1} + R_0 \; ,
\label{eq:poisson}
\end{equation}
in terms of $\bar{n}$, the number of jets expected.  Statistically, a
Poisson distribution represents the number of positive outcomes of
independent trials. We observe it in multiple soft photon radiation
for example off an electron~\cite{peskin}. Its theoretical derivation
rests on two features: first, one splitting dominates, \eg successive
photon radiation off an electron or gluon radiation off a quark;
second, soft radiation is automatically ordered by the radiation
angle, which means there is no combinatorial factor.\medskip

%--------------------------------------------------------
\begin{table}[t]
\begin{tabular}{l|cc}
\hline
& staircase scaling & Poisson scaling \\
\hline
$\sigma_n$ & 
       $\sigma_0 \; e^{-bn}$ & $\hat{\sigma}_0 \; \dfrac{e^{-\bar{n}} \bar{n}^n}{n!}$ \\[5mm] 
$R_{(n+1)/n}$ & $e^{-b}$ & $\dfrac{\bar{n}}{n+1}$ \\[2mm]
$\hat{R}_{(n+1)/n}$ & 
       $e^{-b}$ & $\left( \dfrac{(n+1) \, e^{-\bar{n}} \, \bar{n}^{-(n+1)}}
                                    {\Gamma(n+1) - n \Gamma(n,\bar{n})}
                  + 1 \right)^{-1}$ \\[5mm]
$\langle \nj \rangle$ & $\dfrac{1}{2} \dfrac{1}{\cosh b -1}$ & $\bar{n}$ \\[2mm]
$\pveto$ &
       $1-e^{-b}$ & $e^{-\bar{n}}$ 
\end{tabular}
\caption{Observables for staircase and Poisson scaling. The ratios $R$
  and $\hat{R}$ correspond to the exclusive and inclusive $\nj$
  distribution, respectively. $\Gamma(n,\bar{n})$ is the upper
  incomplete gamma function. In the last line we show the jet veto
  survival probability $\pveto \equiv
  \sigma_0/\hat{\sigma}_\text{tot}$.}
\label{tab:scaling}
\end{table}
%--------------------------------------------------------

From the previous discussion we know that such a radiation pattern
does not occur for total jet rates in proton-proton collisions. The
dominant splitting ISR via the gluon self-coupling instead leads to
staircase scaling.  

This changes once we apply cuts forcing our events into a specific
kinematic configuration. The form of the Poisson distribution is
reminiscent of Sudakov factors, a solution to the DGLAP equation on
which the parton shower approach is based.  Such Sudakov factors model
the leading non-splitting probabilities for example of partons
entering the hard process coming from the proton.  The approximation
underlying the DGLAP equation is collinearity and a sizable
difference in virtualities between partons inside the proton and the
hard process. It then corresponds to resumming the collinear logarithm
arising from successive ISR.

\begin{figure}[t]
\includegraphics[width=0.47\textwidth]{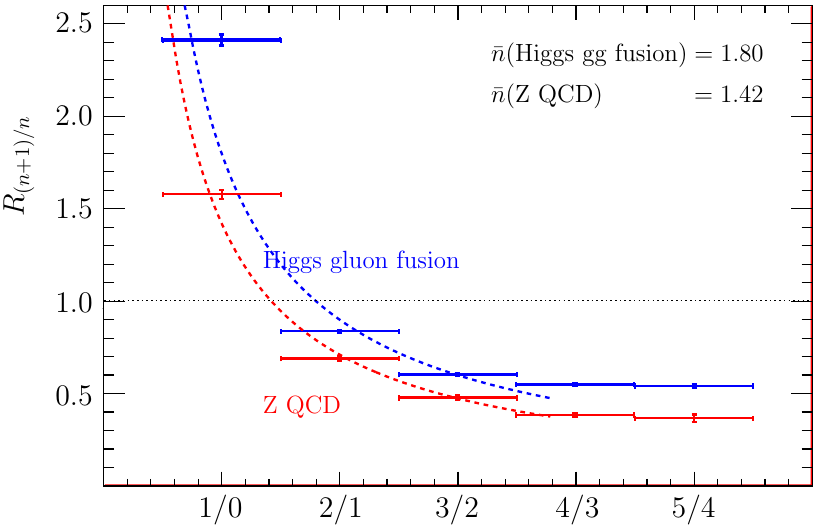}
\caption{$R_{(n+1)/n}$ distributions for $Z$+jets production 
  and Higgs production via the effective gluon-gluon-Higgs
  coupling. Unlike in Fig.~\ref{fig:staircase1} we now count only
  additional jets in the veto region defined by
  Eq.\eqref{eq:veto}. The curves are fits to Eq.\eqref{eq:poisson}.}
\label{fig:poisson}
\end{figure}

A similar kinematic situation we enforce through the WBF cuts
Eq.\eqref{eq:wbf}: the partons have to generate the large invariant
mass $m_{jj} > 600$~GeV which favors quarks in the initial
state. These quarks can most efficiently evolve via successive
soft-collinear gluon emission; \ie they show the same radiation
pattern as soft photons being radiated off a hard electron. The moment
such a large (collinear) logarithm enhances jet radiation we observe
Poisson scaling of the jet ratios in QCD $Z$+jets and $H$+jets, 
Fig.~\ref{fig:poisson}.  Values of $R_{1/0} \gg 1$ reflect a much 
higher probability to observe jet radiation.

Because Poisson scaling only enhances radiation as long as the
large logarithm dominates the kinematics, for large $\nj$ we do not
approach zero, but the staircase limit. Fitting Poisson curves to the first
three bins in Fig.~\ref{fig:poisson} gives us $\bar{n}=1.4$ for QCD
$Z$ production and $\bar{n} = 1.8$ for Higgs production in gluon
fusion. The last two bins follow a staircase pattern.  A comparison
with Fig.~\ref{fig:staircase2} confirms that we see consistent $R$
values in the staircase setup and the high-multiplicity Poisson
regime.\medskip

%--------------------------------------------------------
%--------------------------------------------------------

Comparing the QCD and electroweak $Z$+jets production processes we
observe a very clear difference. While the staircase scaling property
is indeed slightly sculpted due to the different subprocesses, the
Poisson shape for the QCD process significantly enhances the first two
jet radiations.

Among other observables, Tab.~\ref{tab:scaling} tells us how to
compute a jet veto survival probability from the exclusive jet
scaling. Based on our numerical results we cannot expect to simply
extract the two scaling parameters $b$ and $\bar{n}$ and insert them
into the resummed form for $\pveto$. On the other hand, the distinct
patterns shown in Tab.~\ref{tab:scaling} allow us to study the $\nj$
distributions before and after cuts, validate the appropriate
simulation, and quantify the scaling ratios $R_{(n+1)/n}$ including
their uncertainties. For WBF signal-like processes, where a
measurement is difficult, it is crucial that we can understand and
simulate the staircase scaling pattern which is typical for inclusive
LHC processes.\medskip

\underline{Outlook} --- In this Letter we have laid out a strategy to
understand jet counting and the associated veto survival
probabilities. The tool behind this study is the exclusive $\nj$
distribution, as currently measured in several processes at the
LHC.\medskip

There exist two scaling types for exclusive jet rates at the LHC.
First, staircase scaling for inclusive as well as for exclusive jet
rates is defined by a constant ratio $R = \sigma_{n+1}/\sigma_n$.
It occurs for total cross sections for example in $W/Z/\gamma$+jets
and pure QCD jet production~\cite{scaling_orig,us,photon} and is due
to non-Abelian gluon splitting.  In Higgs production staircase scaling
is realized by all signal and background inclusive rates and by the
signal and electroweak $Z$ production after tagging jet cuts.

Second, Poisson scaling appears for processes in which jet radiation
is enhanced by a large logarithm.  It corresponds to successive gauge
boson radiation.  In the past, this scaling has been used to describe
all signal and background processes after tagging jet
cuts~\cite{wbf,cjv}.  For QCD $Z$+jets and gluon-fusion Higgs
production we confirm this result after requiring two tagging jets.

The main effect of these two scaling patterns is visible in the first
emission. While for electroweak processes we find $R_{1/0} =
0.15-0.3$, typical QCD processes after WBF cuts range around $R_{1/0}
\sim 1.5-2.5$. An exclusive jet requirement or central jet veto is a
powerful tool to suppress backgrounds, provided that kinematic cuts
drive the backgrounds into a Poisson regime and leave the signal of
staircase type.

An understood $\nj$ distribution with and without Poisson-inducing
cuts allows us to carefully study all aspects of jet counting, including
experimental and theoretical uncertainties. It should enable LHC searches
which otherwise are plagued by severe theoretical
uncertainties.\medskip

\underline{Acknowledgments} --- We would like to thank Peter 
Schichtel for pointing out an error in the text which misidentified 
the tagging jet selection.  EG would like to thank 
the Institut f\"ur Theoretische Physik in Heidelberg for continuing 
hospitality and always sunny weather. 

%%%%%%%%%%%%%%%%%%%%%%%%%%%%%%%%%%%%%%%%%%%%%%%%%%%%%%%%%%%%%%%%%%%%%%%%


\baselineskip15pt

\begin{thebibliography}{99}


\bibitem{atlas}
 ATLAS~Collaboration, Report No. ATLAS-CONF-2011-111 .

\bibitem{cms}
 CMS~Collaboration, Report No. CMS-HIG-11-003.

\bibitem{wbf}
 D.~L.~Rainwater, D.~Zeppenfeld and K.~Hagiwara,
  %``Searching for H --> tau tau in weak boson fusion at the LHC,''
  Phys.\ Rev.\  D {\bf 59}, 014037 (1999);
  %[arXiv:hep-ph/9808468].
  %%CITATION = PHRVA,D59,014037;%%
 T.~Plehn, D.~L.~Rainwater and D.~Zeppenfeld,
  %``A method for identifying H --> tau tau --> e+- mu-+ missing p(T)  at the
  %CERN LHC,''
  Phys.\ Rev.\  D {\bf 61}, 093005 (2000).
  %[arXiv:hep-ph/9911385].
  %%CITATION = PHRVA,D61,093005;%%

\bibitem{cjv}
 U.~Baur, E.~W.~N.~Glover,
  %``Tagging the Higgs boson in p p ---> W+ W- j j,''
  Phys.\ Lett.\  {\bf B252}, 683-689 (1990);
 V.~D.~Barger, K.~Cheung, T.~Han, D.~Zeppenfeld,
  %``Single n jet tagging and central jet vetoing to identify the leptonic $W W$ decay mode of a heavy Higgs boson,''
  Phys.\ Rev.\  {\bf D44}, 2701 (1991);
 V.~D.~Barger, R.~J.~N.~Phillips, D.~Zeppenfeld,
  %``Mini - jet veto: A Tool for the heavy Higgs search at the LHC,''
  Phys.\ Lett.\  {\bf B346}, 106-114 (1995).
  %[hep-ph/9412276].  
 D.~L.~Rainwater, R.~Szalapski, D.~Zeppenfeld,
  %``Probing color singlet exchange in $Z$ + two jet events at the CERN LHC,''
  Phys.\ Rev.\  {\bf D54}, 6680-6689 (1996);
  %[hep-ph/9605444].
 D.~L.~Rainwater, D.~Summers, D.~Zeppenfeld,
  %``Multi - jet structure of high E($T$) hadronic collisions,''
  Phys.\ Rev.\  {\bf D55}, 5681-5684 (1997).
  %[hep-ph/9612320].
 T.~Figy, V.~Hankele, D.~Zeppenfeld,
  %``Next-to-leading order QCD corrections to Higgs plus three jet production in vector-boson fusion,''
  JHEP {\bf 0802}, 076 (2008).
  %[arXiv:0710.5621 [hep-ph]].  
  %\cite{Cox:2010ug}
  B.~E.~Cox, J.~R.~Forshaw, A.~D.~Pilkington,
  %``Extracting Higgs boson couplings using a jet veto,''
  Phys.\ Lett.\  {\bf B696}, 87-91 (2011).
  %\cite{Figy:2007kv}
    
\bibitem{wh}
  J.~M.~Butterworth, A.~R.~Davison, M.~Rubin, G.~P.~Salam,
  %``Jet substructure as a new Higgs search channel at the LHC,''
  Phys.\ Rev.\ Lett.\  {\bf 100}, 242001 (2008).
  %[arXiv:0802.2470 [hep-ph]].

\bibitem{scet}
  for recent theory developments see \eg
  I.~W.~Stewart, F.~J.~Tackmann,
  %``Theory Uncertainties for Higgs and Other Searches Using Jet Bins,''
  [arXiv:1107.2117 [hep-ph]].

\bibitem{early_matching}
 for an early correct description of one jet over its entire $p_T$ range see:
 M.~Bengtsson, T.~Sjostrand,
  %``Coherent Parton Showers Versus Matrix Elements: Implications of PETRA - PEP Data,''
  Phys.\ Lett.\  {\bf B185}, 435 (1987);
 G.~Miu, T.~Sjostrand,
  %``$W$ production in an improved parton shower approach,''
  Phys.\ Lett.\  {\bf B449}, 313-320 (1999);
  %[hep-ph/9812455].
 M.~H.~Seymour,
  %``A Simple prescription for first order corrections to quark scattering and annihilation processes,''
  Nucl.\ Phys.\  {\bf B436}, 443-460 (1995);
  %[hep-ph/9410244].
 G.~Corcella, M.~H.~Seymour,
  %``Initial state radiation in simulations of vector boson production at hadron colliders,''
  Nucl.\ Phys.\  {\bf B565}, 227-244 (2000).
  %[hep-ph/9908388].

\bibitem{ckkw}
  S.~Catani, F.~Krauss, R.~Kuhn, B.~R.~Webber,
  %``QCD matrix elements + parton showers,''
  JHEP {\bf 0111}, 063 (2001),
  %[hep-ph/0109231].
  S.~H\"oche, F.~Krauss, S.~Schumann and F.~Siegert,
  %  % ``QCD matrix elements and truncated showers,''
   JHEP {\bf 0905}, 053 (2009).
  %[arXiv:0903.1219 [hep-ph]].

\bibitem{mlm}
  M.~L.~Mangano, M.~Moretti, R.~Pittau,
  %``Multijet matrix elements and shower evolution in hadronic collisions: $W b \bar{b}$ + $n$ jets as a case study,''
  Nucl.\ Phys.\  {\bf B632}, 343-362 (2002).
  %[hep-ph/0108069].

\bibitem{us}
 C.~Englert, T.~Plehn, P.~Schichtel, S.~Schumann,
  %``Jets plus Missing Energy with an Autofocus,''
  Phys.\ Rev.\  {\bf D83}, 095009 (2011).
  %[arXiv:1102.4615 [hep-ph]].

\bibitem{ex_lhc}
 ATLAS~Collaboration, G. Aad {\it et al.}
  %``Measurement of the production cross section for W-bosons in association with jets in pp collisions at sqrt(s) = 7 TeV with the ATLAS detector,''
  [arXiv:1012.5382 [hep-ex]],
  %%CITATION = ARXIV:1012.5382;%%
% ATLAS~Collaboration,
  %``Measurement of the production cross section for W- and Z-bosons in association with jets in ATLAS,''
  [arXiv:1106.2061 [hep-ex]],
% ATLAS~Collaboration,
  %``Measurement of dijet production with a veto on additional central jet activity in pp collisions at sqrt(s)=7 TeV using the ATLAS detector,''
  [arXiv:1107.1641 [hep-ex]],
% ATLAS~Collaboration,
  %``Measurement of multi-jet cross sections in proton-proton collisions at a 7 TeV center-of-mass energy,''
  and [arXiv:1107.2092 [hep-ex]];
 
 CMS Collaboration,
  %``Measurement of the Ratio of the 3-jet to 2-jet Cross Sections in pp Collisions at sqrt(s) = 7 TeV,''
  [arXiv:1106.0647 [hep-ex]];
    The D0 Collaboration, V.M. Abazov {\it et al.}, Phys. Lett. B {\bf 682}, 370 (2010);
  The CDF Collaboration, T. Aaltonen {\it et al.}, Phys. Rev. Lett. {\bf 100}, 102001 (2008). 

\bibitem{scaling_orig}
 S.~D.~Ellis, R.~Kleiss, W.~J.~Stirling,
  %``W's, Z's and Jets,''
  Phys.\ Lett.\  {\bf B154}, 435 (1985);
 F.~A.~Berends, W.~T.~Giele, H.~Kuijf, R.~Kleiss, W.~J.~Stirling,
  %``MULTI - JET PRODUCTION IN W, Z EVENTS AT p anti-p COLLIDERS,''
  Phys.\ Lett.\  {\bf B224}, 237 (1989);
 F.~A.~Berends, H.~Kuijf, B.~Tausk, W.~T.~Giele,
  %``On the production of a W and jets at hadron colliders,''
  Nucl.\ Phys.\  {\bf B357}, 32-64 (1991).
  
\bibitem{photon}
  S.~H\"oche, S.~Schumann, F.~Siegert,
  %``Hard photon production and matrix-element parton-shower merging,''
  Phys.\ Rev.\  {\bf D81}, 034026 (2010).
  %[arXiv:0912.3501 [hep-ph]].
 C.~Englert, T.~Plehn, P.~Schichtel, and S.~Schumann,
  in preparation.

\bibitem{LH_2009_ggH}
  J.~M.~Butterworth {\it et al.},
  %``The Tools and Monte Carlo working group Summary Report,''
  [arXiv:1003.1643 [hep-ph]].

\bibitem{ex_tev}
  V.~M.~Abazov {\it et al.} [ D0 Collaboration ],
  %``Measurement of the ratios of the Z/gamma* + >= n jet production cross sections to the total inclusive Z/gamma* cross section in p anti-p collisions at s**(1/2) = 1.96-TeV,''
  Phys.\ Lett.\  {\bf B658 } (2008)  112-119;
  % [hep-ex/0608052].
  %%CITATION = hep-ex/0608052%%
  T.~Aaltonen {\it et al.} [ CDF - Run II Collaboration ],
  % ``Measurement of inclusive jet cross-sections in Z/gamma*(---> e+ e-) + jets production in p anti-p collisions at s**(1/2) = 1.96-TeV,''
  Phys.\ Rev.\ Lett.\  {\bf 100 } (2008)  102001.
  % [arXiv:0711.3717 [hep-ex]].
  %%CITATION = arXiv:0711.3717%%

\bibitem{sherpa}
  T.~Gleisberg, S.~H\"oche, F.~Krauss, M.~Sch\"onherr, S.~Schumann, F.~Siegert, J.~Winter,
  %``Event generation with SHERPA 1.1,''
  JHEP {\bf 0902}, 007 (2009).
  %[arXiv:0811.4622 [hep-ph]].

\bibitem{cteq10}
  H.~-L.~Lai, M.~Guzzi, J.~Huston, Z.~Li, P.~M.~Nadolsky, J.~Pumplin, C.~-P.~Yuan,
  %``New parton distributions for collider physics,''
  Phys.\ Rev.\  {\bf D82}, 074024 (2010).
  %[arXiv:1007.2241 [hep-ph]].

\bibitem{blackhat}
  C.~F.~Berger, Z.~Bern, L.~J.~Dixon {\it et al.},
  %``Precise Predictions for W + 4 Jet Production at the Large Hadron Collider,''
  arXiv:1009.2338 [hep-ph].
  %%CITATION = ARXIV:1009.2338;%%

\bibitem{wbf_nlo}
 T.~Han, G.~Valencia and S.~Willenbrock,
  %``Structure function approach to vector boson scattering in p p collisions,''
  Phys.\ Rev.\ Lett.\  {\bf 69}, 3274 (1992);
  %[arXiv:hep-ph/9206246].
  %%CITATION = PRLTA,69,3274;%%
% M.~Spira,
%  %``QCD effects in Higgs physics,''
%  Fortsch.\ Phys.\  {\bf 46}, 203 (1998);
%  %[arXiv:hep-ph/9705337].
%  %%CITATION = FPYKA,46,203;%%
 T.~Figy, C.~Oleari and D.~Zeppenfeld,
  %``Next-to-leading order jet distributions for Higgs boson production via
  %weak-boson fusion,''
  Phys.\ Rev.\  D {\bf 68}, 073005 (2003);
  %[arXiv:hep-ph/0306109].
  %%CITATION = PHRVA,D68,073005;%%
 M.~Ciccolini, A.~Denner and S.~Dittmaier,
  %``Strong and electroweak corrections to the production of Higgs+2jets via
  %weak interactions at the LHC,''
  Phys.\ Rev.\ Lett.\  {\bf 99}, 161803 (2007)
  %[arXiv:0707.0381 [hep-ph]].
  %%CITATION = PRLTA,99,161803;%%
and 
% M.~Ciccolini, A.~Denner and S.~Dittmaier,
%  %``Electroweak and QCD corrections to Higgs production via vector-boson fusion
%  %at the LHC,''
   Phys.\ Rev.\  D {\bf 77}, 013002 (2008);
%  %[arXiv:0710.4749 [hep-ph]].
%  %%CITATION = PHRVA,D77,013002;%%
 P.~Bolzoni, F.~Maltoni, S.~O.~Moch and M.~Zaro,
  %``Higgs production via vector-boson fusion at NNLO in QCD,''
  Phys.\ Rev.\ Lett.\  {\bf 105}, 011801 (2010).
  %[arXiv:1003.4451 [hep-ph]].
  %%CITATION = PRLTA,105,011801;%%

\bibitem{peskin}
  M.~E.~Peskin, D.~V.~Schroeder,
  %``An Introduction to quantum field theory,''
  Reading, USA: Addison-Wesley (1995) 842 p.
  
\end{thebibliography}
\end{document}